\documentclass[aps,prl,twocolumn,superscriptaddress,oneside,floatfix,amsmath,showpacs,amssymb]{revtex4}
\usepackage{graphicx}
\usepackage{dcolumn}
\usepackage{bm}

\begin{document}

\newcommand{\Y}{YBa$_2$Cu$_3$O$_{6.5}$}
\newcommand{\ie}{\textit{i.e.}}
\newcommand{\eg}{\textit{e.g.}}
\newcommand{\etal}{\textit{et al.}}


\title{Quantum oscillations in underdoped YBa$_2$Cu$_3$O$_{6.5}$}


\author{Cyril~Jaudet}
\affiliation{Laboratoire National des Champs Magn\'{e}tiques
Intenses (CNRS), Toulouse, France}

\author{Julien~Levallois}
\affiliation{Laboratoire National des Champs Magn\'{e}tiques
Intenses (CNRS), Toulouse, France}

\author{Alain~Audouard}
\affiliation{Laboratoire National des Champs Magn\'{e}tiques
Intenses (CNRS), Toulouse, France}

\author{David~Vignolles}
\affiliation{Laboratoire National des Champs Magn\'{e}tiques
Intenses (CNRS), Toulouse, France}

\author{Baptiste~Vignolle}
\affiliation{Laboratoire National des Champs Magn\'{e}tiques
Intenses (CNRS), Toulouse, France}

\author{Ruixing~Liang}
\affiliation{Department of Physics and Astronomy, University of
British Columbia, Vancouver, Canada} \affiliation{Canadian
Institute for Advanced Research, Toronto, Canada}

\author{D.A.~Bonn}
\affiliation{Department of Physics and Astronomy, University of
British Columbia, Vancouver, Canada} \affiliation{Canadian
Institute for Advanced Research, Toronto, Canada}

\author{W.N.~Hardy}
\affiliation{Department of Physics and Astronomy, University of
British Columbia, Vancouver, Canada} \affiliation{Canadian
Institute for Advanced Research, Toronto, Canada}

\author{N.E.~Hussey}
\affiliation{H. H. Wills Physics Laboratory, University of
Bristol, Bristol, U.K.}

\author{Louis~Taillefer}
\affiliation{Canadian Institute for Advanced Research, Toronto,
Canada} \affiliation{D\'epartement de physique \& RQMP,
Universit\'e de Sherbrooke, Sherbrooke, Canada}

\author{Cyril~Proust} \email{proust@lncmp.org}
\affiliation{Laboratoire National des Champs Magn\'{e}tiques
Intenses (CNRS), Toulouse, France} \affiliation{Canadian Institute
for Advanced Research, Toronto, Canada}

\begin{abstract}
Shubnikov-de Haas and de Haas-van Alphen effects have been
measured in the underdoped high temperature superconductor
YBa$_2$Cu$_3$O$_{6.51}$. Data are in agreement with the standard
Lifshitz-Kosevitch theory, which confirms the presence of a
coherent Fermi surface in the ground state of underdoped cuprates.
A low frequency $F = 530 \pm 10$ T is reported in both
measurements, pointing to small Fermi pocket, which corresponds to
$2\%$ of the first Brillouin zone area only. This low value is in
sharp contrast with that of overdoped
Tl$_2$Ba$_2$CuO$_{6+\delta}$, where a high frequency $F = 18$~kT
has been recently reported and corresponds to a large hole
cylinder in agreement with band structure calculations. These
results point to a radical change in the topology of the Fermi
surface on opposing sides of the cuprate phase diagram.
\end{abstract}


\maketitle


The first unambiguous observation of quantum oscillations (QO) in
underdoped YBa$_2$Cu$_3$O$_{6.5}$ \cite{Doiron07} has created a
lot of interest concerning the exact nature and the origin of the
Fermi surface (FS) in the pseudogap phase of cuprates. If QO are
interpreted as the usual consequence of the quantization of closed
orbits in a magnetic field, they show that the FS consists of
small pockets. It contrasts with angle-resolved photoemission
spectroscopy (ARPES) which shows an apparent destruction of the FS
producing a set of disconnected Fermi arcs \cite{Norman98}. The
apparent contradiction between the two sets of measurements is not
yet resolved \cite{Chakravarty03, Harrison07, Hossain08}.
Shubnikov-de Haas effect has been confirmed in the stoichiometric
compound YBa$_2$Cu$_4$O$_8$ \cite{Bangura07, Yelland07},
indicating that QO are generic to the CuO$_2$ planes rather than
some feature of the band structure specific to
YBa$_2$Cu$_3$O$_{6.5}$ \cite{Bangura07,Carrington07}. More
recently, de Haas-van Alphen effect (dHvA) has been reported in
underdoped YBa$_2$Cu$_3$O$_{6.5}$ \cite{Jaudet08, Sebastian08},
which confirmed the existence of a well-defined, closed, and
coherent, FS via a thermodynamic probe.

In this paper we report a set of new data of Shubnikov-de Haas
(SdH) in an underdoped cuprate YBa$_2$Cu$_3$O$_{6.51}$ obtained
both in the in-plane and Hall resistances. We compare the results
with measurements of the dHvA at the same doping level published
in ref.~\cite{Jaudet08}.

Transverse magnetoresistance ($R_{xx}$) and Hall resistance
($R_{xy}$) were measured down to $1.5$ K in pulsed magnetic fields
up to 60~T. Current and field were applied along the \textbf{a}
axis and normal to the $CuO_2$ planes, respectively. A commercial
piezoresistive microcantilever \cite{Ohmichi02} was used for
torque measurements in a dilution fridge down to ~$0.5$~K in
pulsed magnetic fields up to 59~T. The crystal was glued to the
beam of the cantilever with an angle of approximately $\theta
\thicksim 5^o$ between the direction of the magnetic field and the
normal to the $CuO_2$ planes \cite{Jaudet08}.

\begin{figure}
\centering
\includegraphics[width=0.85\linewidth,angle=0,clip]{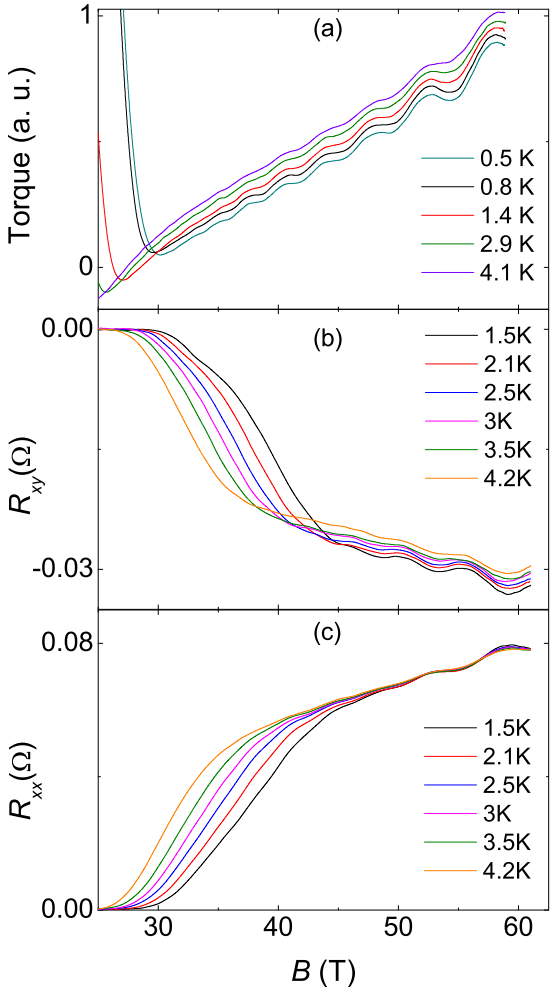}
\caption{Field dependence of the torque (a), Hall resistance (b)
and magnetoresistance (c) at different temperatures. The curves of
the torque are shifted for clarity.} \label{Fig1}
\end{figure}

We used two detwinned single crystals of YBa$_2$Cu$_3$O$_{6.51}$
flux-grown in a non-reactive BaZrO$_3$ crucible \cite{Liang00}.
The dopant oxygen atoms are ordered into an ortho-II
superstructure of alternating full and empty CuO chains which
corresponds to a superconducting temperature of $T_c = 57.5$~K and
therefore a doping of about $p \sim 0.1$\cite{Liang06}.

Fig.~\ref{Fig1}(a) shows raw data of torque (defined as
$\tau=|\overrightarrow{M}\times\overrightarrow{B}|$ where $M$ is
the magnetization) in one sample of YBa$_2$Cu$_3$O$_{6.51}$ at
different temperatures. The torque signal displays clear dHvA
oscillations in high magnetic fields. Above $30$ T, the torque
varies almost linearly with magnetic field which allow us to
subtract a monotonic background from the data in order to derive
the oscillatory part. Eight oscillations can be resolved at
$T=0.5$ K.

Fig.~\ref{Fig1}(b) and Fig.~\ref{Fig1}(c) show raw data of the
Hall resistance and of the transverse magnetoresistance,
respectively, in another sample of YBa$_2$Cu$_3$O$_{6.51}$ at
different temperatures. Oscillatory parts of $R_{xy}$ and $R_{xx}$
are obtained after substraction of a polynomial background.

\begin{figure}
\centering
\includegraphics[width=0.85\linewidth,angle=0,clip]{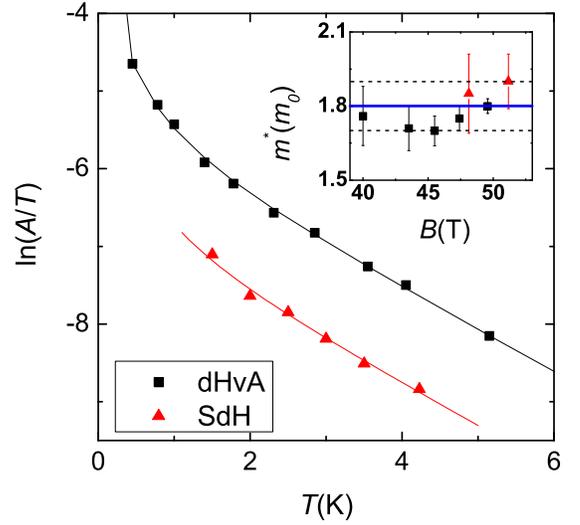}
\caption{Temperature dependence of the amplitude of the
oscillations A of the dHvA and the SdH oscillations at a mean
magnetic field of 47.4~T and 51.7~T, respectively. Solid lines are
best fit to eq.~\ref{equation1}. Inset: Field dependence of the
effective mass.} \label{Fig2}
\end{figure}

The oscillatory part of the resistance ($\Delta R$) and of the
magnetization ($\Delta M$) are analyzed in the framework of the
standard Lifshitz-Kosevich (LK) theory for a two-dimensional
generalized Fermi liquid \cite{Schoenberg84}:

\begin{equation}
\label{equation1} \Delta R, \Delta M \propto R_T R_D cos \Big[2\pi
\big( \frac{F}{B}-\gamma \big) \Big]
\end{equation}


where $F$ is the oscillation frequency and $\gamma$ a phase
factor. The amplitude is given by $A \propto R_T R_D$, where $R_T
= \alpha T m^*/B$ sinh$[\alpha T m^*/B]$ and $R_D = $exp$[-\alpha
T_D m^*/B]$ are the thermal and Dingle damping factors,
respectively. $m^*$ is the cyclotron effective mass, $T_D$ is the
Dingle temperature ($T_D \propto \tau^{-1}$) and $\alpha = 2\pi^2
k_B m_0 / e \hbar$ ($\sim 14.69$ T/K).

Fourier analysis of the oscillatory part of the QO all give a
single peak at about $F = 530 \pm 10$ T, which is also consistent
with the previous study \cite{Doiron07}. The amplitude of the
oscillations in $R_{xx}$ is large enough in the present study to
deduce reliable parameters from the Fourier analysis. In
particular, it is clear from Fig.~\ref{Fig1} that the phase shift
between the transverse magnetoresistance $R_{xx}$ and the Hall
resistance $R_{xy}$ is about $\pi$, as expected in the case of an
electron pocket. Such a pocket in the FS of underdoped
YBa$_2$Cu$_3$O$_{y}$ has been inferred from Hall effect
measurements \cite{LeBoeuf07} and confirmed in the present study
in a different sample.

Fig.~\ref{Fig2} displays the temperature dependence of the
oscillation amplitudes. Good agreement with the LK theory is
observed, giving an effective mass $m^*=1.8 \pm 0.1~m_0$. The
inset of Fig.~\ref{Fig2} shows the effective mass deduced from
dHvA and SdH in different field windows. Within our experimental
resolution, we can conclude that the effective mass is
field-independent, as expected in the standard LK theory
\cite{Schoenberg84}.

\begin{figure}
\centering
\includegraphics[width=0.85\linewidth,angle=0,clip]{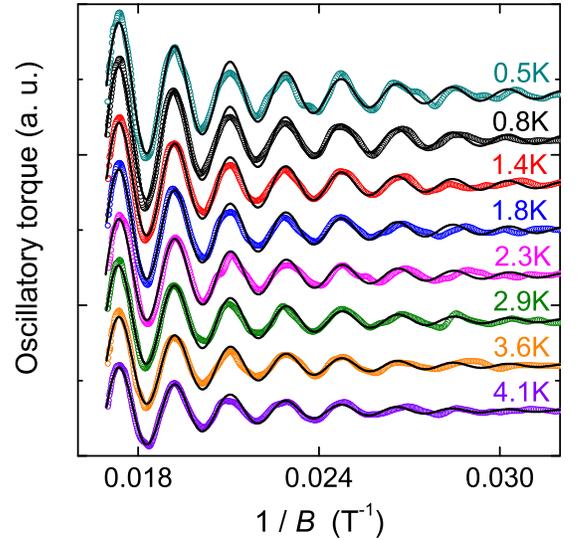}
\caption{Oscillatory part of the torque as a function of the
inverse field for various temperatures. Solids lines are best fit
to eq.~\ref{equation1} with $m^*=1.76~m_0$ and $T_D= 6.2$
K.}\label{Fig3}
\end{figure}

The field dependence of the amplitude of the oscillations leads to
a Dingle temperature $T_D=6.2 \pm 1.2$~K for dHvA data and
converts to a mean free path of $ l\sim160~\AA$ and $w_c \tau =
0.7 \pm 0.2$ at 35~T. Within our experimental resolution, it is
not possible to resolve an extra attenuation corresponding to the
effect of superconductivity on the dHvA amplitude. Note that the
field range in which $R_{xy}$ oscillations are detected is too
small to derive a reliable Dingle temperature.

Fig.~\ref{Fig3} displays the oscillatory torque versus 1/$B$
between $T$ = 4.1~K and $T$ = 0.5~K. Black solid lines are fits to
Eq.~\ref{equation1} obtained by setting $m^*$ = 1.76~$m_0$ and
$T_D$ = 6.2~K for all temperatures. The deduced oscillation
frequency and phase factor are $F$ = 540 $\pm$ 4~T and $\gamma$ =
0.15 $\pm$~0.05, respectively. This fitting procedure confirms
that the LK theory, which describes the dHvA oscillations for a
generalized 2D Fermi liquid, is appropriate.

\begin{figure}
\centering
\includegraphics[width=0.85\linewidth,angle=0,clip]{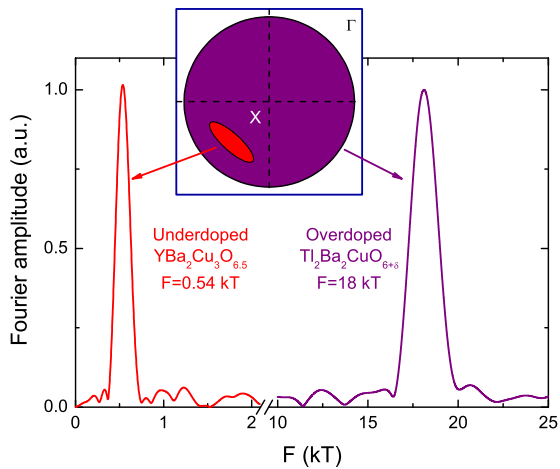}
\caption{Fourier transform of the oscillatory part of the torque
for underdoped $YBa_2Cu_3O_{6.51}$ (red) and for overdoped Tl2201
(purple). Insert: Sketch of the FS area deduced from the
frequencies of quantum oscillations via the Onsager relation. Note
that the location in k-space and the number of small red pocket is
at present unknown.} \label{Fig4}
\end{figure}

The angular dependence of the SdH frequency \cite{Jaudet08}
indicates that QO arise from a quasi-2D FS. The Onsager relation
$F = (\phi_0/2\pi^2)A_k $ ($\phi_0$ is the flux quantum) yields a
cross sectional area $A_k = 5.1$ nm$^{-2}$ which corresponds to a
small pocket covering $\sim 1.9 \%$ of the first Brillouin zone
(FBZ). In overdoped Tl$_2$Ba$_2$CuO$_{6+\delta}$ (Tl2201)
superconductor, AMRO \cite{Hussey03}, ARPES \cite{Plate05} and
more recently QO \cite{Vignolle08} measurements indicate that the
FS consists of a quasi-2D hole cylinder of $\sim 65\%$ of the FBZ
area, in agreement with LDA calculation \cite{Singh92}. The sharp
contrast between the size of the FS on opposite sides of the phase
diagram is illustrated in Fig.~\ref{Fig4}. The main panel displays
the Fourier transform of QO in underdoped YBa$_2$Cu$_3$O$_{6.51}$
(red) and of overdoped Tl2201 (purple)\cite{Vignolle08}. The inset
shows the area of the corresponding orbit in the FBZ. This drastic
dissimilarity of FS topology simply reflects the difference in
carrier density on opposite sides of the phase diagram. While the
large cylinder in the overdoped side corresponds to a carrier
density of $1 + p$, as predicted by band structure calculations,
the band picture fails in the underdoped side, where the carrier
number scales more closely with $p$. Other indications pointing to
a low carrier concentration in underdoped cuprates come from
thermodynamic \cite{Timusk00}, transport\cite{Hussey08} and
superfluid density\cite{Broun07}.

To be more precise, the carrier density is given by the Luttinger
theorem, $n=A_k/(2\pi^2)=F/\phi_0$ for each pocket. The frequency
$F$ = 530~T converts to a carrier density of 0.038 carriers per
planar Cu atom. Independently of whether there are 2 or 4 pockets,
it gives a number of carriers which is not in agreement with the
doping level (10~$\%$). However, a scenario based on a
reconstruction of the FS can explain both the negative Hall effect
(electron pocket) in the normal state \cite{LeBoeuf07} and the
apparent violation of the Luttinger sum rule. It assumes that the
frequency observed with SdH and dHvA effects corresponds to an
electron pocket, whose mobility is much higher than that of a
larger hole pocket. Note that this larger pocket is not detected
in the present measurement, but may have been detected in a recent
dHvA study in the same sample \cite{Sebastian08}. The origin of
such reconstruction is still under debate: a reconstruction could
occur due to a competing order parameter in the pseudogap phase,
e.g. antiferromagnetism \cite{Chen08}, d-DW
\cite{Dimov08,Podolsky08} or orbital currents\cite{Zhu08}, or a
stripe-like order could appear close to the 1/8 doping
\cite{Millis07}.

In summary, we have measured the SdH and dHvA effects in
underdoped YBa$_2$Cu$_3$O$_{6.51}$ in pulsed magnetic fields up to
60~T. A single frequency $F = 530 \pm 10$ T has been reported in
both measurements. The data are well described by the LK theory,
which accounts for the dHvA oscillations of generalized 2D Fermi
liquid. The low frequency corresponds to a small pocket covering
$\sim 1.9 \%$ of the FBZ, in sharp contrast with the large
cylinder predicted by band structure calculations and observed
experimentally in overdoped Tl$_2$Ba$_2$CuO$_{6+\delta}$. The
fundamental question is: What causes the FS to undergo such a
radical change between the overdoped and the underdoped regions of
the phase diagram?

\end{document}